\documentclass[a4paper,11pt]{article}
\pdfoutput=1 

\usepackage{jheppub} 

\usepackage[T1]{fontenc} 

\title{\boldmath The possibility of leptonic CP-violation measurement with JUNO }

\author[a,1]{M.V. Smirnov,\note{Corresponding author.}}
\author[a]{Zh.J. Hu,}
\author[a]{S.J. Li}
\author[a,1]{{\rm and} J.J. Ling}

\affiliation[a]{Department of physics, Sun Yat-Sen University,\\Guangzhou 510275, China}

\emailAdd{gear8mike@gmail.com}
\emailAdd{lingjj5@mail.sysu.edu.cn}

\abstract{The existence of CP-violation in the leptonic sector is one of the most important issues for modern science. Neutrino physics is a key to the solution of this problem. JUNO (under construction) is the near future of neutrino physics. However CP-violation is not a priority for the current scientific program. We estimate the capability of $\delta_{\rm CP}$ measurement, assuming a combination of the JUNO detector and a superconductive cyclotron as the antineutrino source. This method of measuring CP-violation is an alternative to conventional beam experiments. A significance level of 3$\sigma$ can be reached for 22\% of the $\delta_{\rm CP}$ range. The accuracy of measurement lies between 8$^{\rm o}$ and 22$^{\rm o}$. It is shown that the dominant influence on the result is the uncertainty in the mixing angle $\Theta_{23}$.}

\begin{document} 
\maketitle
\flushbottom

\section{Introduction}
\label{sec:intro}

\par
\indent 

The absence of anti-matter in our Universe is a mystery for modern theoretical physics.
Many different scientific hypotheses and theories were developed for a description of the solution to this issue~\cite{BAS}.   
This asymmetry may be a consequence of the Sakharov conditions being satisfied~\cite{Sakharov}.  
One of which is the breaking of fundamental symmetry between particles and antiparticles, so called CP-violation (CPV). 
In general, CPV can be represented as:\\
\begin{equation}
\label{eq_1}
P(A{\to}B) \ne P(\bar A{\to}\bar B).
\end{equation}
Equation \eqref{eq_1} states that the probabilities for particles and antiparticles in symmetric processes are different.
In 1964 the first evidence of CPV was observed in the quark sector in decays of neutral K-mesons~\cite{CP}. 
Clearer confirmation of CPV was found in the decay of B-mesons~\cite{CP2}. Such process can be expressed in terms of CKM-mixing\footnote{The Cabibbo-Kobayashi-Maskawa mixing matrix for quarks} matrix. Unfortunately observed CPV in the quark sector is quite small and can not explain the current matter-antimatter asymmetry.

The leptonic sector is also a promising space within which to search for CPV. After experimental confirmation of  neutrino oscillation~\cite{osc_obs}, it became clear that neutrinos have mass. 
Furthermore, the measurement of non-zero mixing angle $\theta_{13}$~\cite{theta_13} opens the door for the observation of CPV in the leptonic sector using neutrinos.

The theory of neutrino oscillation tells us, that the phase of CPV can be  observed only when one neutrino flavor converts to another neutrino flavor, wherein both flavors are known.
 In this paper we assume, that neutrinos are Dirac particles. The most convenient flavors of neutrino for  experimental research are electron and muon neutrinos (antineutrinos).   
Using the standard parameterization of the PMNS\footnote{The Pontecorvo-Maki-Nakagawa-Saka neutrino mixing matrix} mixing matrix, the transition probability between muon and electron flavors of neutrino in vacuum can be written as follows:
\begin{equation}
\begin{split}
\label{eq_2}
P\big(^{{\bar{\nu }}_{\mu }{\to }{\bar{\nu }}_{e}}_{\nu _{\mu }{\to }\nu _{e}}\big)&=\sin ^{2}\theta _{23}\sin ^{2}2\theta _{13}\sin ^{2}\Delta _{31}+
\cos ^{2}\theta _{23}\sin ^{2}2\theta _{12}\sin ^{2}\Delta _{21}+\\
&+\sin 2\theta _{13}\sin 2\theta _{23}\sin 2\theta _{12}\sin \Delta _{31}\sin \Delta _{21}\cdot \cos (\Delta _{31}\mp \delta _{\mathrm{CP}}),
\end{split}
\end{equation}
where $\Delta_{ij}=\Delta m_{ij}^2\cdot L/(4E_{\nu})$; $\Delta m_{ij}^2$ -- the neutrino mass squared difference; $L$ -- the distance between source and detector; $E_{\nu}$ -- neutrino energy.
The term responsible for CPV is included as an argument of the cosine function. 
As can be seen from equation~\eqref{eq_2}, if $\delta_{\rm CP}$ equals 0 or $\pi$, there is no violation of CP-symmetry.

\section{A nonstandard method for measuring CPV}
\par
\indent 

The traditional approach to placing limits on $\delta_{\rm CP}$ is based on the comparison of transition probabilities $P(\nu_{\mu}\to\nu_e)$ and $P(\bar{\nu}_{\mu}\to\bar{\nu}_e)$ or vice versa.
This method allows us to observe the breaking of CP-symmetry directly. 
As a general rule,  CPV experiments  use a powerful beam with a couple of neutrino detectors. One is a near detector, and another is a far detector. The near detector is usually relatively small, and  it is used to measure neutrino flux from the source with minimal oscillation probability. The far detector should be located at  the oscillation maximum for given energy of neutrino, where the splitting for different values of $\delta_{\rm CP}$ is higher. The most common materials for the target of the detector are water and liquid argon. The ability to reconstruct direction using these materials helps greatly in suppressing background. Liquid scintillator (LSc) can also be used in neutrino beam experiments.
The main experiments investigating CPV at this moment are LBNF plus DUNE~\cite{DUNE}; J-PARC plus HyperK~\cite{HyperK}; NuMI plus NO$\nu$A~\cite{NOvA}.

In addition, there are other nonstandard methods of measuring CPV, which are based on using superconducting cyclotrons, high intensity beta-beams~\cite{Agarwalla:2009em}.

\subsection{DAE$\delta$ALUS as a neutrino source}
\par
\indent 

\begin{figure}[ht]
\centering \includegraphics[scale=0.6]{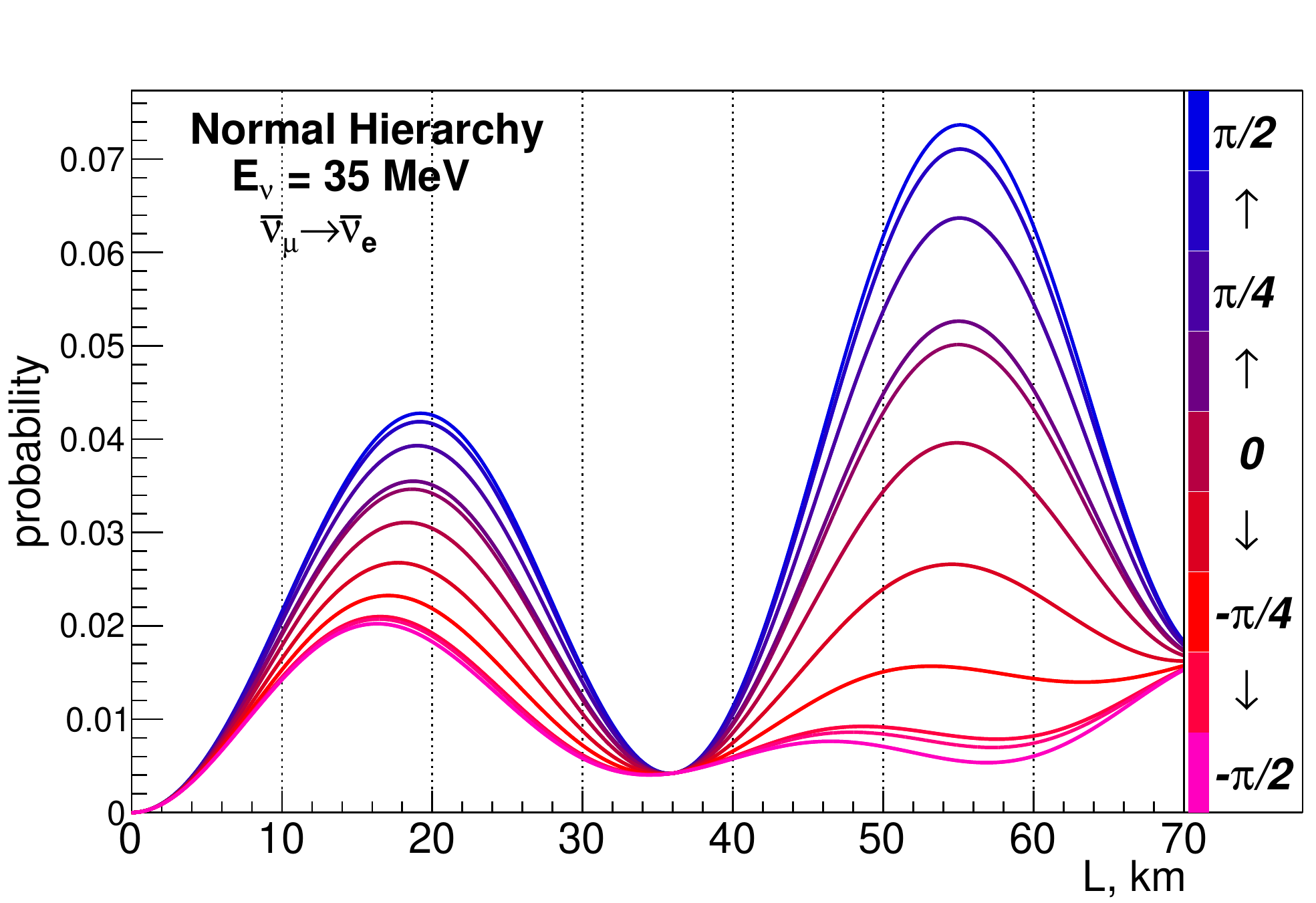}
\caption{\label{fig:1}{The transition probability between $\bar{\nu}_\mu$ and $\bar{\nu}_e$ as a function of the distance L for different values $\delta_{\rm CP}$, for fixed neutrino energy 35 MeV.}}
\end{figure}

The initial proposal for an experiment to probe $\delta_{\rm CP}$ was made in 2010~\cite{daedalus,conrad}.  
The proposed experiment consisted of three superconductive cyclotrons, which are located at 1.5 km (near), 8 km (middle), 20 km (far) with a single Water-Cherenkov (WC) detector of total mass 300 kt. However, there were other proposals. For instance, to use a LSc detector (LENA)~\cite{Wurm}. The expected energy of the cyclotrons proton beam is 650--1500 MeV/n. The power of a single cyclotron should equal 1 MW. The proton beam will hit the graphite target and produce $\pi^{\pm}$.
$\pi^{-}$ will be quickly captured by the surrounding matter. 
After that the stopped $\pi^{+}$ will decay at rest to $\mu^{+}$ and $\nu_{\mu}$, then $\mu^{+}\to e^+\bar{\nu}_\mu\nu_e$. The experiment is focused on the transition from muon antineutrino to electron antineutrino.
Only one oscillation channel is considered in comparison with beam experiments,  which may use four oscillation transitions ($\stackrel{{(-)}}{\nu}_{\mu}\iff\stackrel{(-)}{\nu}_{e}$).
It should be mentioned, that such a measurement is insensitive to the mass hierarchy (MH), while the traditional method requires that we know the MH beforehand.
The behavior of the oscillation curve is shown in figure~\ref{fig:1}. The influence of $\delta_{\rm CP}$ on the oscillation curve is obvious. The first oscillation maximum is located $\approx$20 km. The energy spectrum of muon antineutrinos is continuous with a 52.8 MeV endpoint. The total power of neutrino sources should equal 1 MW, 2 MW, 5 MW, for near, middle and far cyclotrons respectively. The expected running time is 10 years with a duty factor of 20\%.
The main detection channel is inverse beta-decay (IBD). This explains the choice of WC detector, although LSc detector can be used for such an experiment as well. A good approximation for the IBD cross-section is~\cite{IBD_xsec}:
\begin{equation}
\label{eq_3}
\sigma_{\rm IBD}\approx p_eE_eE_{\nu}^{-0.07056+0.02018\ln E_\nu-0.001953\ln^3E_\nu}\cdot10^{-43}{\rm cm^2},
\end{equation}
where $p_e$ -- momentum of the positron, $E_e$ -- energy of the positron, $E_\nu$ -- antineutrino energy. 
The expected event rate in the energy window 20--52.8 MeV for $\delta_{\rm CP}=\pi/2$ should equal 1600 events for 10 years of measurements. The background is negligible in this energy region, and the most of it comes from atmospheric neutrinos. Signal events exceed the background fourfold for the maximal possible event rate in a 300 kt WC detector.    

\section{JUNO and modified DAE$\delta$ALUS}
\subsection{Liquid scintillator detector JUNO}
\par
\indent 

The Jiangmen Underground Neutrino Observatory \\
(JUNO) is located in Guangdong province, China~\cite{juno}. The 20 kt detector consists of two nested acrylic spheres placed within a high purity water pool. The inner sphere has a radius of 17.7 m, and the outer sphere radius is 20 m. The high purity of the LSc allows to achieve a new extremely low value for the energy resolution $3\%/\sqrt{E({\rm MeV})}$. 
The main goal of JUNO is the determination of the mass hierarchy.
However, as shown in this paper, this neutrino detector can also be used to measure or place limits on CPV.
	
\subsection{Proposal}
\par
\indent

\begin{figure}[ht]
\centering \includegraphics[scale=0.45]{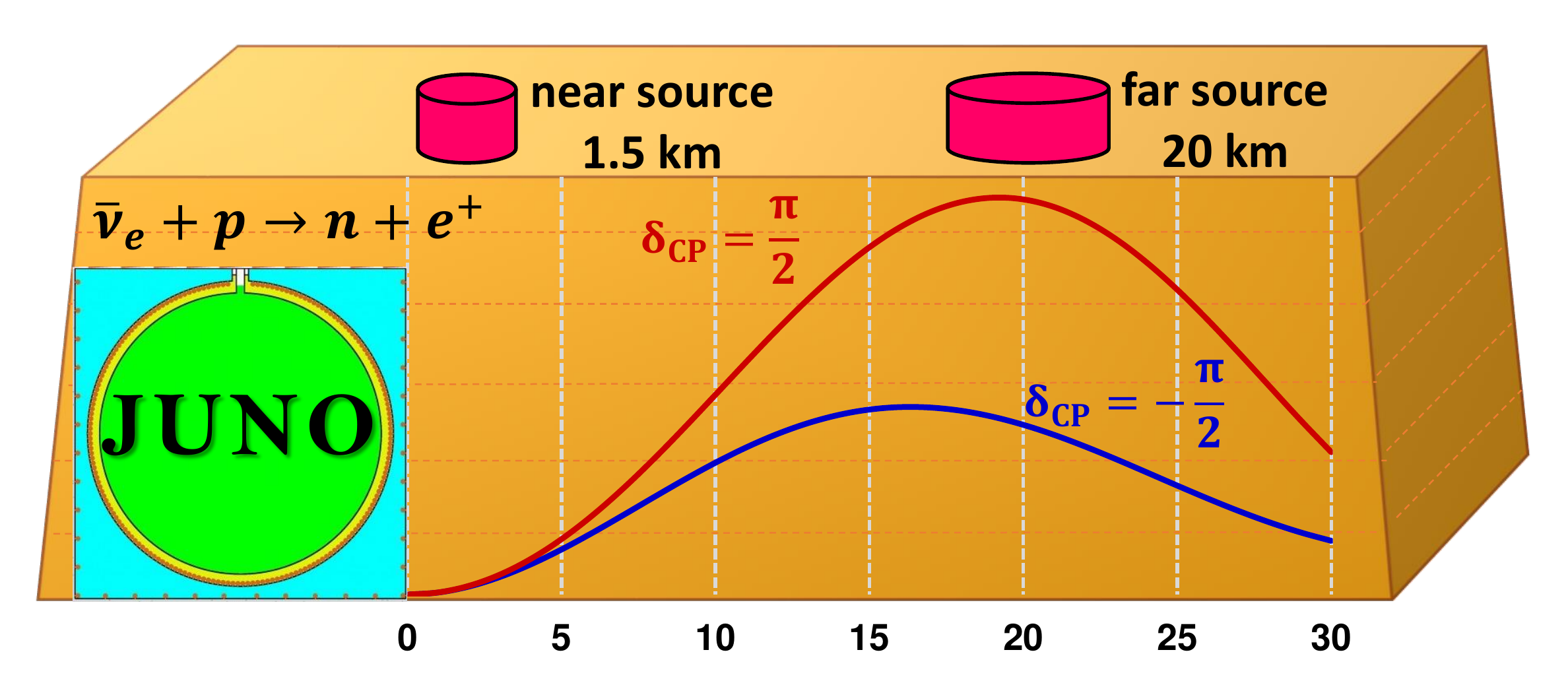}
\caption{\label{fig:2} Schematic layout of the experiment with two neutrino sources (near and far) and the JUNO detector.}
\end{figure}

Our proposal consists of the combination of two projects, JUNO as a detector and DAE$\delta$ALUS as  a neutrino source. Additionally we suggest a modification of the neutrino source configuration.
Instead of using three cyclotrons in different locations,  we consider only two cyclotrons. One near and another far at distances 1.5 km and 20 km respectively. The approximate scheme of the experiment is shown in figure~\ref{fig:2}. 

\begin{figure}[ht]
\centering \includegraphics[scale=0.35]{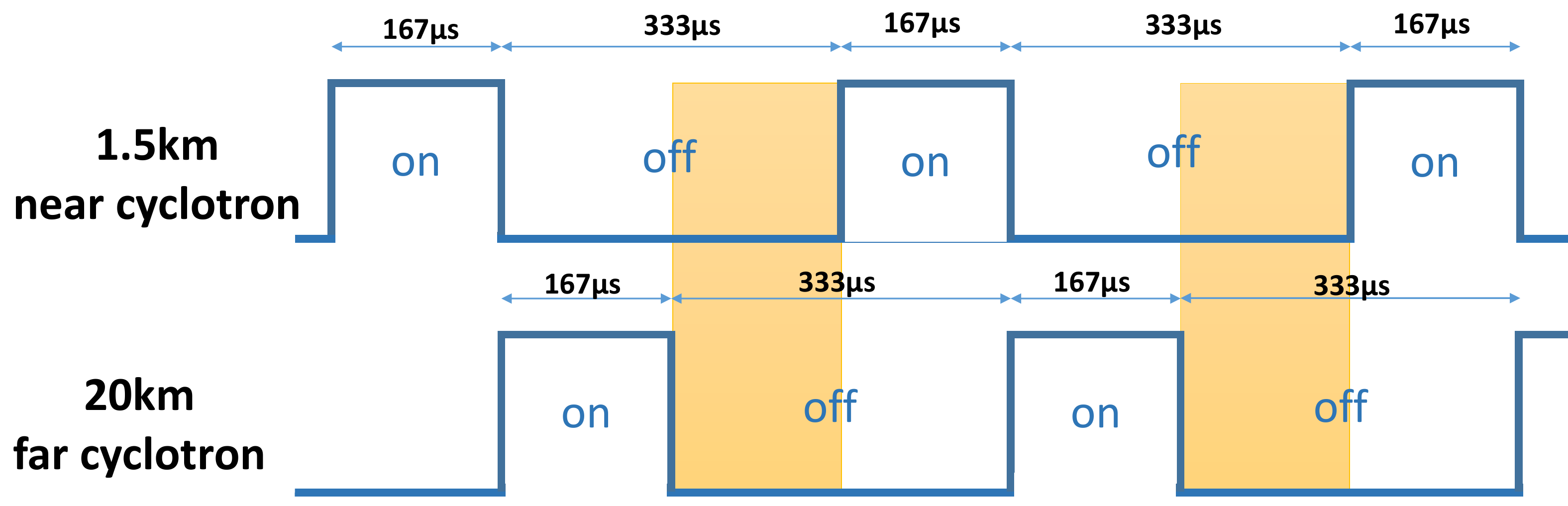}
\caption{\label{fig:3} The running time of each cyclotron. The yellow rectangles show the beam-off time, which will be used for background measurements.}
\end{figure}

The power of the near cyclotron is 1 MW. It is needed for flux normalization and near-physics experiments~\cite{daedalus}. For the far cyclotron there are two options. 
One is a standard power  cyclotron of 5 MW\footnote{The original proposal by DAE$\delta$ALUS group}, and another is increased to 10 MW. Since the middle cyclotron is not included, it is reasonable to increase the working time of each cyclotron to 33\% of exposure time. This is illustrated by a pulse diagram in figure~\ref{fig:3}. In principle the duty factor can be very close to 100\% for single cyclotron~\cite{Evslin:2015pya}.
The target material for pion production is identical to DAE$\delta$ALUS (graphite), with yield of pions 0.172 per proton. The expected running time of the proposed experiment is 10 years.

\subsection{Event rate analysis}
\par
\indent

\begin{figure}[ht]
\centering
\begin{minipage}[b]{0.45\linewidth}
\centering
\includegraphics[scale=0.405]{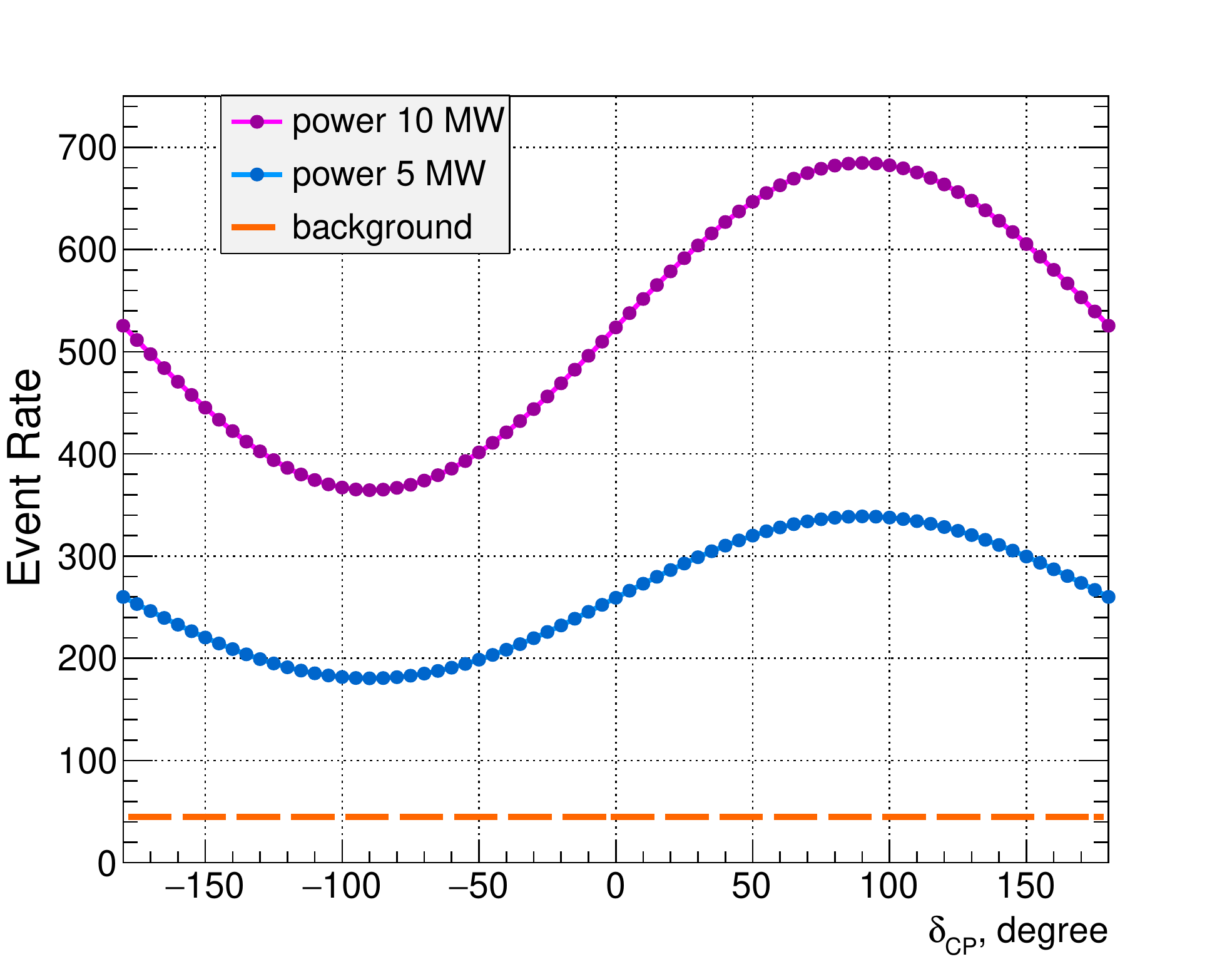}
\end{minipage}
\qquad
\begin{minipage}[b]{0.49\linewidth}
\centering
\includegraphics[scale=0.405]{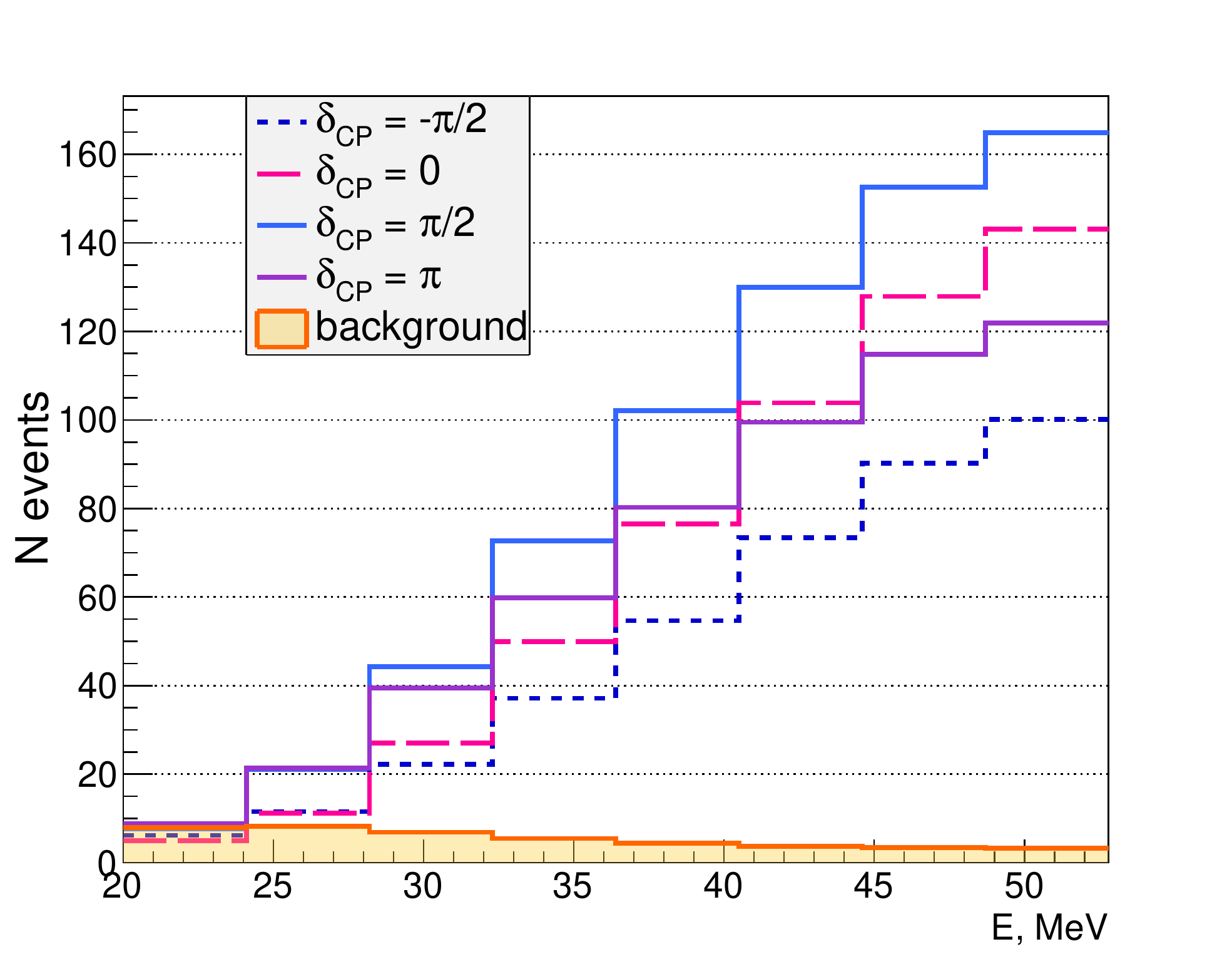} 
\end{minipage}
\caption{\label{fig:4}The left panel shows the event rate for JUNO as a function of $\delta_{\rm CP}$ at the distance 20 km for two different powers of far  cyclotron 5 MW and 10 MW. 
The right panel shows the shape of the IBD events spectrum for four different values $\delta_{\rm CP}$ (assumed power of the cyclotron is 10 MW and an exposure time 200 kt$\cdot$year), and the shape of the background.}
\end{figure}

As we consider only the IBD reaction as the main channel for detecting neutrino events, the event rate can be estimated by using
equation~\eqref{eq_4}:
\begin{equation}
\label{eq_4}
dN=\Phi(L)\cdot T\cdot n_p\cdot\sigma(E_\nu)\cdot P(L, E_\nu)\cdot S(E_\nu)dE_\nu,
\end{equation}
where $\Phi(L)$ is neutrino flux at the distance $L$; $T$ -- exposure time; $n_p$ -- number of free protons in the volume of the detector\footnote{The number of free protons is larger for LSc than for water}; $\sigma(E_\nu)$ -- IBD cross-section \eqref{eq_3}; $P(L,E_\nu)$ -- oscillation probability function for transition $\bar{\nu}_{\mu}\to\bar{\nu}_e$ \eqref{eq_2}; $S(E_\nu)$ -- the shape of neutrino spectrum, in our case it is the Michel spectrum of $\bar{\nu}_{\mu}$. Based on equation \eqref{eq_4} and using the latest values of oscillation parameters from PDG~\cite{PDG}. The estimated event rate and the shape of the antineutrino spectrum is illustrated in figure~\ref{fig:4}.
As can be seen in figure~\ref{fig:4}, the maximal event rate corresponds to a CP phase of $\pi/2$, whereas the minimal value is $-\pi/2$. 

This statistical analysis includes two parts. The first part is the sensitivity of the experiment to CPV. 
The second part is the accuracy with which $\delta_{\rm CP}$ can be measured.

\subsubsection{\label{sec_3_3_1}Experimental sensitivity to discovery of CPV}
The most standard approach to the quantification of sensitivity to CPV is the minimization of a $\Delta\chi^2$ function~\cite{delta_chi,Ge:2012wj}.
This function can be written as:
\begin{equation}
\label{eq_5}
\Delta\chi^2_{\rm CPV}={\rm\bf min}[\chi^2(\delta_{\rm CP}^{\rm test}=0|\pi)-\chi^2(\delta_{\rm CP}^{\rm true})].
\end{equation}
$\chi^2(\delta_{\rm CP}^{\rm test}=0|\pi)$  refers to  two different chi-square functions. One is for a fixed value of CP phase equal to 0, another is for a fixed value $\pi$. 
Then we minimize $\chi^2(\delta_{\rm CP}^{\rm test}=0|\pi)$ considering both cases and 
whichever gives the lowest value is plugged back into~\eqref{eq_5}. 
Significance level  can be defined as $\sigma=\sqrt{\Delta\chi^2_{\rm CPV}}$.
The $\Delta\chi^2$ function should have approximately Gaussian distribution with mean value $\Delta\chi^2_{\rm CPV}$ and  standard deviation $2\sqrt{\Delta\chi^2_{\rm CPV}}$~\cite{delta_chi}. The width of this distribution gives information about 68\%, 95\% and etc. bands.

\subsubsection{The accuracy of CP phase measurement}
The determination of the accuracy of measurement for particular value of $\delta_{\rm CP}$ can be calculated by minimizing the chi-square function which in this case is the likelihood function below:
\begin{equation}
\label{eq_6}
\chi_{\rm CP}^2=2\sum_{i=1}^{N_b}\Big{[}\mu_i^{\rm min}(\Omega)-n_i+n_i\cdot\ln\frac{n_i}{\mu_i^{\rm min}(\Omega)}\Big{]}+\sum_{j=1}^{N_p}\frac{(\eta_j-\eta_j^o)^2}{(\delta\eta_j)^2},
\end{equation}
where $N_b$ -- total number of bins in histogram; $\Omega$ is a set of parameters including CP phase and oscillation parameters $\eta_j$; $\mu_i^{\rm min}$ -- predicted counts in $i$-th bin; $n_i$ -- observed counts in $i$-th bin (usually experimental data or MC events); $N_p$ -- amount of oscillation parameters; $\eta_j^o$ -- best fit value of $\eta_j$ (usually  from PDG); $\delta\eta_j$ -- one sigma error of $\eta_j^o$.
\begin{table}[tbp]
\centering
\renewcommand{\arraystretch}{1.1}
\begin{tabular}{lccccc}
\hline
\hline
$\eta_j$ & $\Delta m_{21}^2\cdot10^{-5}~{\rm eV}^2$ & $\Delta m_{32}^2\cdot10^{-3}~{\rm eV}^2$ &$\sin^2(\Theta_{12})$ & $\sin^2(\Theta_{23})$ & $\sin^2(\Theta_{13})\cdot10^{-2}$ \\
\hline 
\hline
$\eta_j^o$ & 7.53 & 2.45 & 0.307 & 0.51 & 2.10\\
$\delta\eta_j$ & 0.18 & 0.05 & 0.013 & 0.04 & 0.11\\
\hline
\end{tabular}
\caption{\label{tab_1} The list of oscillation parameters and their uncertainties from PDG, which were used for the calculation of systematic effects.}
\end{table}
All values of oscillation parameters are presented in  table~\ref{tab_1}.

\subsubsection{Monte-Carlo simulations}
The observed antineutrino spectrum was built on the basis of  equation~\eqref{eq_4} using MC methods. In this analysis, we consider two  types of uncertainties: statistical fluctuations and systematic effects related to oscillation parameters. 
Also the energy resolution of JUNO was added to this analysis.
The main source of background is atmospheric neutral current (NC) events\footnote{Most of them are reactions on carbon}. Our estimation gives 439 NC events for an exposure time of 200 kt$\cdot$year with duty factor 33\%.
However, this background can be significantly decreased as demonstrated in~\cite{Randolph}. 
Searching for  a coincident signal from the decay of a final isotope can reduce background by 40\%.
Using pulse shape discrimination with an acceptance level of 95\%, the background can be decreased eight-fold. 
After these manipulations the total amount of NC events is 33. 
Including atmospheric charge current (CC) and fast neutron events, the total background is 45 events.
Moreover the experiment uses beam-off--beam-on measurements, which may help in the background subtraction. All these reasons allow us to neglect the influence of background on the result. 
Following the original DAE$\delta$ALUS paper ~\cite{daedalus}, which showed the insensitivity of $\delta_{\rm CP}$ measurement to systematic effects associated with flux normalization. 
Thus we do not account for these effects in this analysis.
To estimate sensitivity to CPV, 3.5K MC simulations were calculated for each sample with particular values of $\delta_{\rm CP}$. Both  chi-squares in  the function~\eqref{eq_5} were minimized using the ROOT package Minuit ~\cite{minuit,minuit2}.
Finally, the sensitivity curve was calculated with $\sigma$-level, which was defined in the section~\ref{sec_3_3_1}. 
$\sigma$-level was calculated as the square root of the mean value of the Gaussian distribution for $\Delta\chi_{\rm CP}^2$. 

In order to determine the accuracy of a potential $\delta_{\rm CP}$ measurement 10K MC simulations were calculated for each sample with particular value of $\delta_{\rm CP}$. The chi-square function~\eqref{eq_6} was minimized and the value of $\delta_{\rm CP}^{\rm fit}$ was extracted. $\delta_{\rm CP}^{\rm fit}$ should have Gaussian distribution with mean value $\delta_{\rm CP}^{\rm true}$. The standard deviation of this distribution gives us 1$\sigma$ error for each concrete value of CP phase.

\section{Results} 
\par
\indent

\begin{figure}[ht]
\centering
\begin{minipage}[b]{0.44\linewidth}
\centering
\includegraphics[scale=0.335]{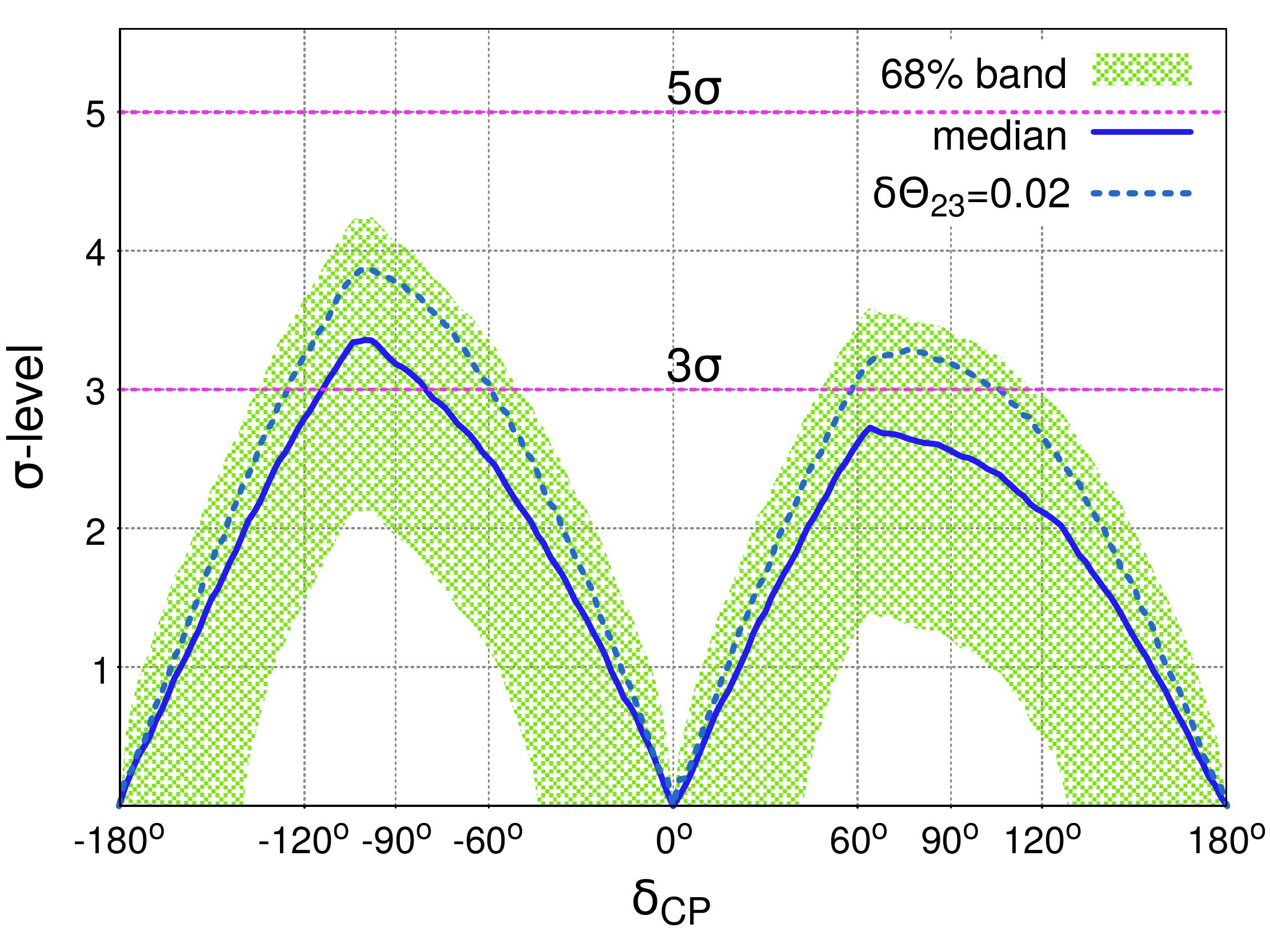}
\end{minipage}
\qquad
\begin{minipage}[b]{0.495\linewidth}
\centering
\includegraphics[scale=0.335]{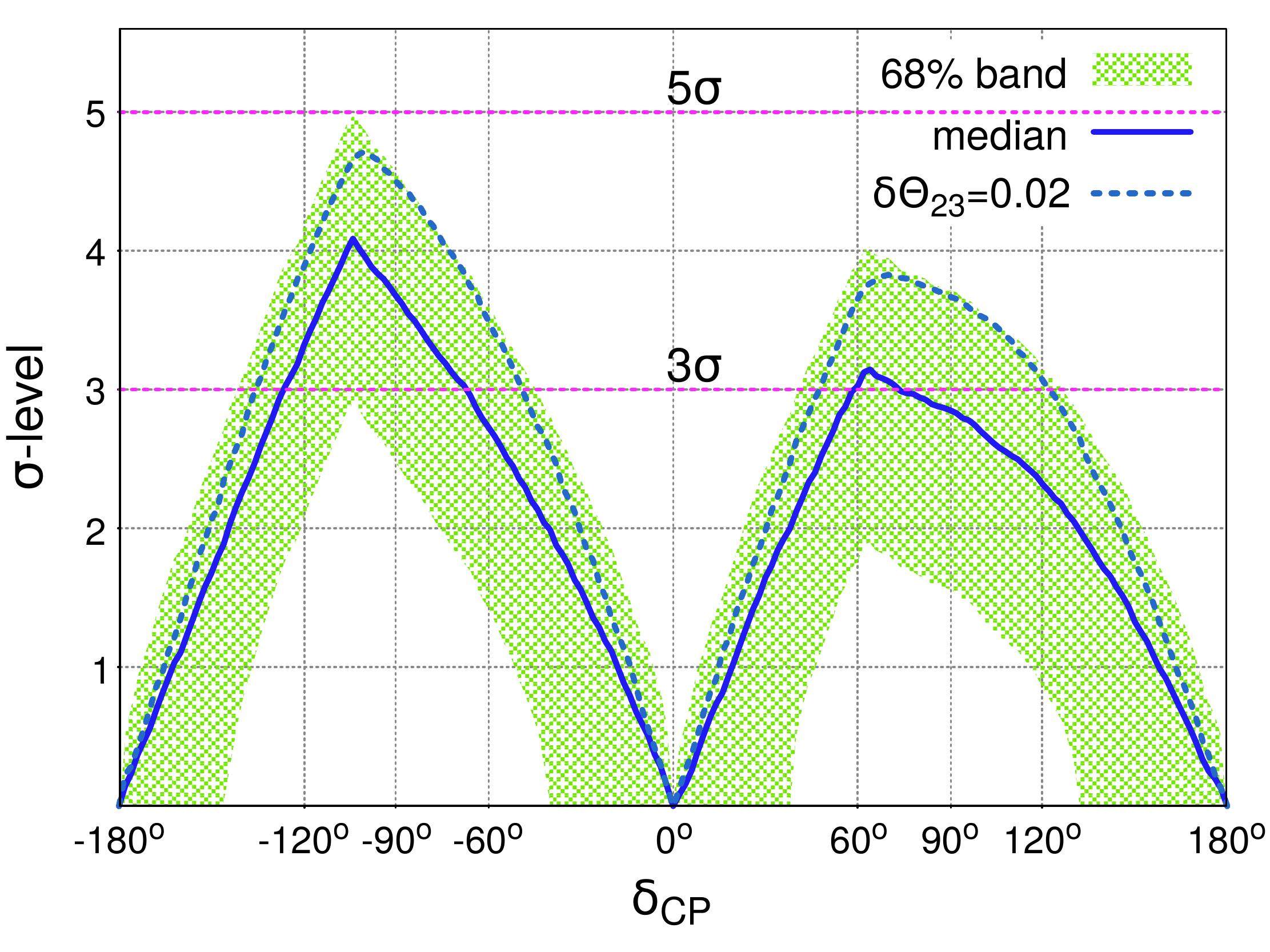} 
\end{minipage}
\caption{\label{fig:5}The significance level for the determination of CPV as a function of $\delta_{CP}$ assuming an exposure time of 200 kt$\cdot$year. The left panel corresponds to 5 MW source power, the right panel -- 10 MW. Sigma level is defined as $\sigma=\sqrt{\Delta\chi^2_{\rm CPV}}$. The dashed line corresponds to the sensitivity, when absolute error for $\sin^2(\Theta_{23})$ is 0.02.}
\end{figure}

Experimental sensitivity to CPV for JUNO is shown on figure~\ref{fig:5}. The green band shows the 68\% confidence interval,  based on statistical and systematic fluctuations.
A 5 MW cyclotron can cover only 10\% of the $\delta_{\rm CP}$ range with a significance level of 3$\sigma$. 	
Whereas a 10 MW cyclotron can reach 22\% of the $\delta_{\rm CP}$ range with a significance level of 3$\sigma$.
 Assuming a twofold decrease of the uncertainty in the mixing angle $\Theta_{23}$, sensitivity can be increased vastly up to 33\% and 45\% for 5 MW and 10 MW respectively.
The asymmetry in each figure can be explained by the presence of $\cos(\delta_{\rm CP})$ in equation~\eqref{eq_2}.

\begin{figure}[ht]
\centering \includegraphics[scale=0.7]{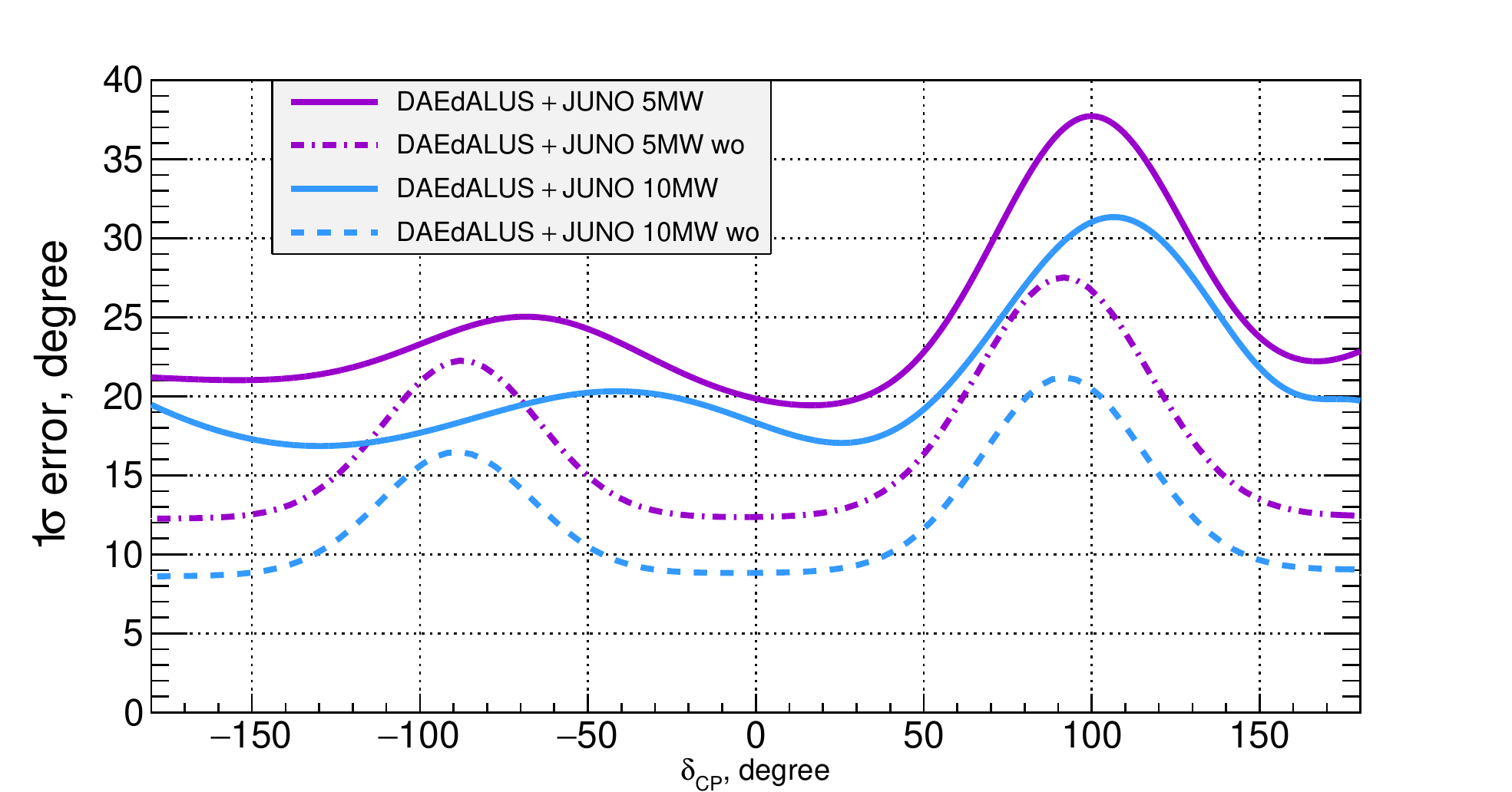}
\caption{\label{fig:6} The accuracy of measurement as a function of $\delta_{\rm CP}$. The solid line corresponds to the case with systematic effects. The dashed line is without systematic effects.}
\end{figure}

Figure~\ref{fig:6} depicts the behavior of the uncertainty for each concrete value of $\delta_{\rm CP}$ .  
Two cases are considered, with systematic effects, and without. 
As can be seen in figure \ref{fig:6}, systematic and statistical effects have  influence on the final result.
Two peaks near $\pm90^o$ confirm the features of the uncertainty of $\delta_{\rm CP}$ given in ~\cite{Coloma:2012wq}.
Our estimation shows, that the dominant contribution to systematic uncertainty comes from the mixing angle $\Theta_{23}$.

\section{Summary and discussions}
\par
\indent

From this research, it follows that not only beam experiments can be used for CPV measurements. Superconductive cyclotrons are another opportunity for investigating CPV. 
It was shown that significance level 3$\sigma$ can be reached and the error of $\delta_{\rm CP}$ lies between $\rm 8^o$ and $\rm 22^o$ for the best case, assuming the uncertainties of oscillation parameters are tiny. Future neutrino experiments will decrease these uncertainties,  especially the most important oscillation parameter $\Theta_{23}$ for sensitivity to CPV.   
At the same time, LSc is a good option for the measuring IBD events in the energy window 20--52.8 MeV. This channel is not available for liquid argon detectors. 
Unfortunately, the prices of superconductive cyclotrons are still high.
However, large scale neutrinos experiments such as JUNO, should explore all possibilities for measuring CPV. Thus far the only proposal  for measuring $\delta_{\rm CP}$ phase utilizes atmospheric neutrinos. 
LSc can not be used effectively for beam experiments, therefore only cyclotrons can provide adequate measurements of CPV in JUNO.

\acknowledgments

This work was in part supported by the National Natural Science Foundation of China (NSFC) under Grant No. 11775315. 
We would like to say great thanks to School of Physics, Sun Yat-Sen University,
especially to leader of our neutrino group Prof. Wei Wang for cultivating good working conditions.
We express special gratitude to Dr. Neill Raper for excellent editing of this paper.


\end{document}